# Controller independent Bidirectional Quantum Direct Communication

Amit Kumar Mohapatra and S. Balakrishnan

**Abstract**  Recently, C. H. Chang *et. al.* (Quan. Info. Proc. 14:3515-3522, 2015) proposed a controlled bidirectional quantum direct communication protocol using four Bell state. In this work, the significance of Bell states, which are being as initial states in C. H. Chang *et. al.* protocol, is elucidated. The possibility for preparing initial state based on the secret message of the communicants is explored. In doing so, the controller independent bidirectional quantum direct communication protocol has evolved naturally. It is shown that any communicant cannot read the secret message without knowing the initial states generated by the other communicant. Further, intercept-and-resend attack and information leakage can be avoided. The proposed protocol is like a conversion between two persons with high level secrecy without the help of any third person.

Amit Kumar Mohapatra and S. Balakrishnan*
Department of Physics
School of Advanced Sciences
VIT University,
Vellore 632014, India
e-mail:omamit1199@gmail.com
*Corresponding author e-mail: physicsbalki@gmail.com

# 1 Introduction

Since the ancient times secret way of communication between the sender and the receiver is an essential task due to various reasons. In the recent years, quantum mechanics is being exploited for the purpose secure communication and hence the subject of quantum cryptography has evolved naturally. In quantum cryptography, generation and distribution of quantum key are extensively used for secure communication between the sender and the receiver. The first quantum key distribution (QKD) protocol is presented by Charles Bennett and Gilles Brassard in the year 1984 [1]. QKD aims at establishing an unconditional secure secret key between two authorized users. Differing from QKD, quantum direct communication (QDC) allows the users to communicate the secret messages directly without creating a key to encrypt them in advance. Since the inception of Ping-Pong protocol by Bostrom and Felbinger [2] in 2002, many QDC protocols have been put forwarded [3-7].

While QDC protocols permit one-way communication between the users, bidirectional QDC (BQDC) protocols allow two users to exchange their secret messages simultaneously. BQSDC is introduced in the year 2004 by Nguyen [4]. Since then the variety of proposals has been suggested for the improvement. In 2005, Man et. al. [5] proposal resolves the problem of intercept-and-resend attack and many schemes have been proposed to overcome the information leakage [8, 10]. Using GHZ state, a controlled bidirectional protocol, in which the exchange of secret messages achieved with a set of device under the control of a third party, is proposed in 2006 [6]. Later, controlled bidirectional quantum direct communication (CBQDC) protocol is proposed using a Bell state instead of GHZ state [9]. However, this protocol due to Ye Tian and Jiang [9] suffers from the information leakage problem and intercept-and-resend attack. A user can obtain the other user's secret message without the controller's permission by performing the intercept-and-resend attack. In order to resolve this problem, C.H. Chang *et. al.* [11] proposed an improvement by using four Bell states as the initial states of controller's resource. In this case, without knowing the initial states the secret messages cannot be communicated. Hence, both intercept-and-resend attack and information leakage are solved.

As the Bell states are the essential physical resource and difficult to implement, it is necessary to find the optimal value of entanglement of quantum states necessary for the execution of C.H. Chang *et. al.* protocol. However, it is found that Bell states are necessary to run the protocol. Further, we present a protocol in which communicants can share the secret messages without the help of the controller. In other words, we exhibit the case of bidirectional quantum direct communication without the influence of the controller. Advantages of the protocol are discussed in the conclusion.

## 2 C.H. Chang et al. protocol

In this section, we brief the C.H. Chang et al. protocol [11], which is an improved version of Ye et al. [9]. The steps adopted in the protocol are given below and the same can be seen in Fig. 1.

**Step 1:** Charlie prepares $(n + l + d)$ Bell states, where each state is in one of the four Bell states:

$$|\phi^+\rangle = \tfrac{1}{\sqrt{2}}(|00\rangle_{AB} + |11\rangle_{AB}), \qquad |\phi^-\rangle = \tfrac{1}{\sqrt{2}}(|00\rangle_{AB} - |11\rangle_{AB}),$$

$$|\psi^+\rangle = \tfrac{1}{\sqrt{2}}(|01\rangle_{AB} + |10\rangle_{AB}) \quad \text{or} \quad |\psi^-\rangle = \tfrac{1}{\sqrt{2}}(|01\rangle_{AB} - |10\rangle_{AB}). \qquad (1)$$

$l$ and $d$ are the numbers for the first and second security checking, $n$ is the number of qubits which will be encoded as a secret message and $\frac{n}{2} = m_1 = m_2$ as suggested by C.H. Chang et al. The subscript $A$ and $B$ denote the first and the second particles of each Bell state that belong to Alice and Bob respectively. Now, Charlie will take the first particles of Bell states to form a sequence $A = [P_1(A), P_2(A), \ldots, P_{n+l+d}(A)]$ and the second particles to form a sequence $B = [P_1(B), P_2(B), \ldots, P_{n+l+d}(B)]$. He will send the sequence $A$ to Alice. For instance, $A_{(n+l+d)} = |\phi^+\rangle = \tfrac{1}{\sqrt{2}}(|00\rangle + |11\rangle)]$ is the initial state prepared by Charlie and the sequences are $A_n = |0\rangle, |1\rangle$ and $B_n = |0\rangle, |1\rangle$. Now, Charlie will send $A_n = |0\rangle, |1\rangle$ sequence to Alice.

**Step 2:** Alice will send a confirmation to Charlie after receiving the sequence $A_n$. Alice will execute the first security checking with Charlie. If error rate goes beyond the threshold, they will abort the communication, else they will continue to the next step.

**Step 3:** Charlie will send the sequence $B_n$ to Bob. Upon receiving the sequence $B_n$, Bob will send a confirmation to Charlie and Alice. Then he will execute the second security checking with Alice.

**Step 4:** If there is no eavesdropper, Alice and Bob will perform the unitary operations according to their secret messages on the sequences $A_n$ and $B_n$ to form the new sequences $A'_n$ and $B'_n$. Usually two-bit secret messages {00, 01, 10, 11} are encoded by the respective unitary operators $\{I, \sigma_z, \sigma_x, i\sigma_y\}$. For secure consideration, Alice and Bob will prepare a sufficient number of single particles $D$ in one of the four states $\{|0\rangle, |1\rangle, |+\rangle, |-\rangle\}$, where $|+\rangle = \tfrac{1}{\sqrt{2}}(|0\rangle + |1\rangle)$ and $|-\rangle = \tfrac{1}{\sqrt{2}}(|0\rangle - |1\rangle)$. By inserting $D$ to the initial state $A'_n$ and $B'_n$, it will become $A'_D$ and $B'_D$ respectively. Alice and Bob then exchange the new states $A'_D$ and $B'_D$ with each other simultaneously and execute the security checking on $D$.

Let Alice's secret message be 10 and Bob's secret message be 01, then, $Msg_A = |10\rangle \to \sigma_x$ and $Msg_B = |01\rangle \to \sigma_z$. After the unitary operation, the new sequences will be $A'_n = \sigma_x(|0\rangle, |1\rangle) = (|1\rangle, |0\rangle)$ and $B'_n = \sigma_z(|0\rangle, |1\rangle) = (|0\rangle, -|1\rangle)$.

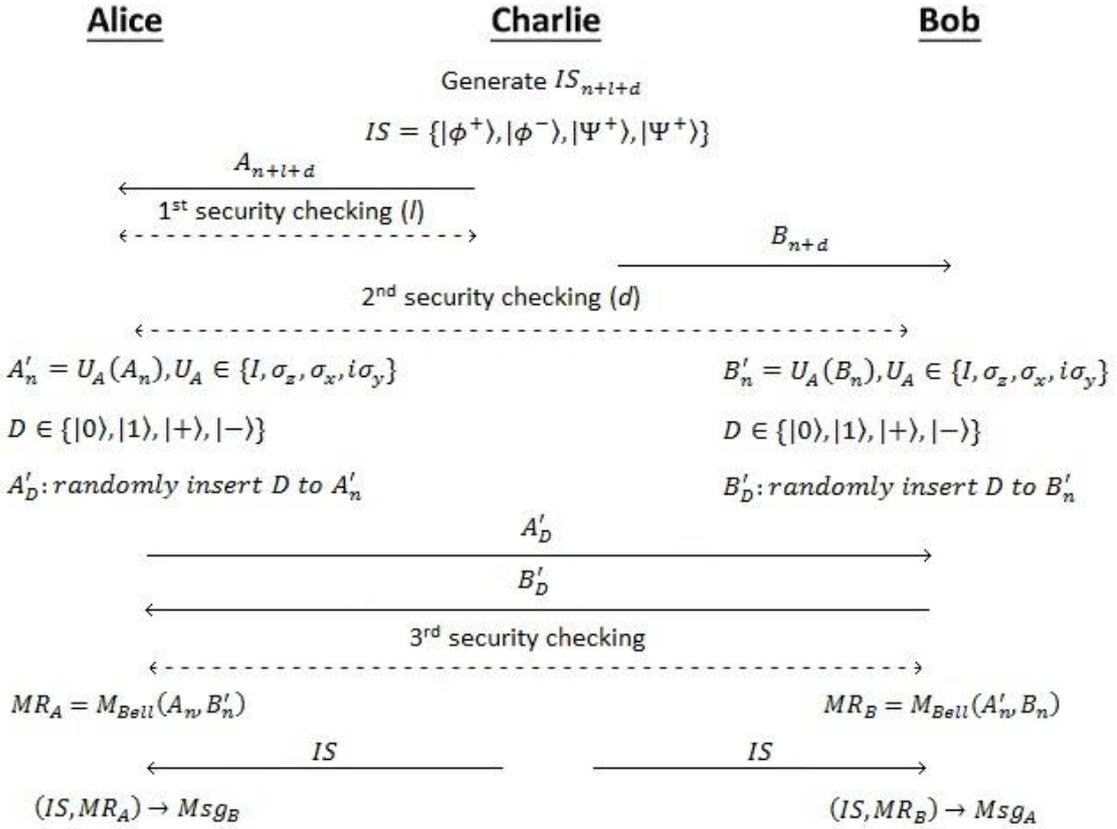

*Fig.1. Chang et al CBQDC protocol*

**Step 5:** Alice and Bob will perform a Bell measurement on the corresponding particles in sequences $B'_n$ and $A'_n$ and obtain the measurement result $MR_A$ and $MR_B$ respectively. If there is no eavesdropper then Charlie will announce the initial state to let Alice and Bob communicate with each other. After getting the initial state and measurement results, Alice and Bob can deduce the secret messages according to Table 1.

By performing Bell measurement on $(A_n, B'_n)$, Alice will get $MR_A = |\phi^-\rangle$ and measurement on $(A'_n, B_n)$ will give $MR_B = |\Psi^+\rangle$ to Bob. After the announcement of initial state by Charlie, Alice will get the Bob's secret message $(IS, MR_A) \to Msg_B = (|\phi^+\rangle, |\phi^-\rangle) \to |01\rangle$ and Bob will get Alice's secret message $(IS, MR_B) \to Msg_A = (|\phi^+\rangle, |\Psi^+\rangle) \to |10\rangle$.

***Table 1***: The relation between Alice's (Bob's) measurements result $MR_A$ ($MR_B$), Charlie's initial states *IS* and Bob's (Alice's) secret messages.

| $MR_A$ ($MR_B$) | | Bob's (Alice's) secret messages | | | |
|---|---|---|---|---|---|
| | | 00 | 10 | 11 | 01 |
| Charlie's *IS* | $\|\phi^+\rangle$ | $\|\phi^+\rangle$ | $\|\psi^+\rangle$ | $\|\psi^-\rangle$ | $\|\phi^-\rangle$ |
| | $\|\psi^+\rangle$ | $\|\psi^+\rangle$ | $\|\phi^+\rangle$ | $\|\phi^-\rangle$ | $\|\psi^-\rangle$ |
| | $\|\psi^-\rangle$ | $\|\psi^-\rangle$ | $\|\phi^-\rangle$ | $\|\phi^+\rangle$ | $\|\psi^+\rangle$ |
| | $\|\phi^-\rangle$ | $\|\phi^-\rangle$ | $\|\psi^-\rangle$ | $\|\psi^+\rangle$ | $\|\phi^+\rangle$ |

As the four initial states, namely Bell states are being controlled by Charlie, the communicants cannot exchange their secret messages without knowing the initial state. Therefore, the intercept and resend attack by any malicious user, say Bob. Further, no information leakage is possible between Alice and Bob in the improved version of Ye et al protocol [9].

## 3 Significance of initial states

In order to exhibit the significance of maximally entangled initial states, the following simple analysis is performed. Instead of four Bell states, we have introduced the following four arbitrary states as Charlie's initial state in the Chang et al protocol:

$$|\omega^+\rangle = \alpha|00\rangle + \beta|11\rangle, \qquad |\omega^-\rangle = \alpha|00\rangle - \beta|11\rangle$$

$$|\chi^+\rangle = \alpha|01\rangle + \beta|10\rangle, \qquad |\chi^-\rangle = \alpha|01\rangle - \beta|10\rangle \qquad (2)$$

where $|\alpha|^2 + |\beta|^2 = 1$. Following the same steps of the Chang protocol as described in the preceding section, we have generated a new table indicating the relation between Alice's (Bob's) measurement result $MR_A$ ($MR_B$), Charlie's initial states *IS* and Bob's (Alice's) secret messages [refer Table 2].

Even if the initial states are non-maximally entangled, Alice and Bob can communicate the secret message 00 or 01. However, the secret messages 10 and 01 cannot be communicated as the measurement results of Alice and Bob are not the same. Therefore, protocol can be executed if and only if both measurement results for the secret messages 01 and 11 are equal. By doing so, we find that $\alpha = \beta = \frac{1}{\sqrt{2}}$, which is turned out be the Bell states as given in Eqn.(1).

***Table 2.*** The relation between Alice's (Bob's) measurement result $MR_A$ ($MR_B$), Charlie's initial states $IS$ and Bob's (Alice's) secret messages.

| $MR_A$ ($MR_B$) | Bob's (Alice's) secret messages | | | |
|---|---|---|---|---|
| | 00 | 10 | 11 | 01 |
| $\lvert\omega^+\rangle = \alpha\lvert 00\rangle + \beta\lvert 11\rangle$ | $\lvert\omega^+\rangle$ ($\lvert\omega^+\rangle$) | $\lvert\chi^+\rangle$ ($\alpha\lvert 10\rangle + \beta\lvert 01\rangle$) | $-\lvert\chi^-\rangle$ ($-\alpha\lvert 10\rangle + \beta\lvert 01\rangle$) | $\lvert\omega^-\rangle$ ($\lvert\omega^-\rangle$) |
| $\lvert\chi^+\rangle = \alpha\lvert 01\rangle + \beta\lvert 10\rangle$ | $\lvert\chi^+\rangle$ ($\lvert\chi^+\rangle$) | $\lvert\omega^+\rangle$ ($\alpha\lvert 11\rangle + \beta\lvert 00\rangle$) | $\lvert\omega^-\rangle$ ($-\alpha\lvert 11\rangle + \beta\lvert 00\rangle$) | $-\lvert\chi^-\rangle$ ($\lvert\chi^-\rangle$) |
| $\lvert\chi^-\rangle = \alpha\lvert 01\rangle - \beta\lvert 10\rangle$ | $\lvert\chi^-\rangle$ ($\lvert\chi^-\rangle$) | $\lvert\omega^-\rangle$ ($\alpha\lvert 11\rangle - \beta\lvert 00\rangle$) | $\lvert\omega^+\rangle$ ($-\alpha\lvert 11\rangle - \beta\lvert 00\rangle$) | $-\lvert\chi^+\rangle$ ($\lvert\chi^+\rangle$) |
| $\lvert\omega^-\rangle = \alpha\lvert 00\rangle - \beta\lvert 11\rangle$ | $\lvert\omega^-\rangle$ ($\lvert\omega^-\rangle$) | $\lvert\chi^-\rangle$ ($\alpha\lvert 10\rangle - \beta\lvert 01\rangle$) | $-\lvert\chi^+\rangle$ ($-\alpha\lvert 10\rangle - \beta\lvert 01\rangle$) | $\lvert\omega^+\rangle$ ($\lvert\omega^+\rangle$) |

It is clear from the analysis that Alice and Bob should able to choose their secret messages irrespective of the initial states generated by Charlie. In other words, secret messages of the communicants cannot be chosen by the initial states. However, it is possible to select the initial states based on the secret messages. This is possible if and only if the communicants have the power to generate the maximally entangled initial states. By doing so, the role of the controller becomes insignificant and we can have controller independent bidirectional direct quantum communication. Having realized these points, the following protocol is being devised.

**4 Controller independent BQDC protocol**

In this new protocol, Alice and Bob are capable of generating Bell states as their initial states. Here we describe how Alice and Bob can exchange their secret messages without depending on the controller, Charlie. The steps followed in the protocol are given in Figure 2.

**Step 1:** Alice prepares Bell states, where each state is one of four bell states $\lvert\phi^+\rangle = \frac{1}{\sqrt{2}}(\lvert 00\rangle_{AB} + \lvert 11\rangle_{AB})$, $\lvert\phi^-\rangle = \frac{1}{\sqrt{2}}(\lvert 00\rangle_{AB} - \lvert 11\rangle_{AB})$, $\lvert\psi^+\rangle = \frac{1}{\sqrt{2}}(\lvert 01\rangle_{AB} + \lvert 10\rangle_{AB})$ or $\lvert\psi^-\rangle = \frac{1}{\sqrt{2}}(\lvert 01\rangle_{AB} - \lvert 10\rangle_{AB})$. The first particles of the Bell states belong to Alice and the second particles belong to Bob. Firstly Alice will choose a Bell state randomly and the unitary operator according to the secret messages. Two-bit secret messages {00, 01, 10, 11} encoded as unitary operators { $I, \sigma_z, \sigma_x, i\sigma_y$ } respectively. Now Alice will perform the unitary operations according to her secret messages on the Bell state and then she will announce the result to Bob and Charlie.

Say, $A$ is the initial state $[A = |\phi^+\rangle = \frac{1}{\sqrt{2}}(|00\rangle + |11\rangle)]$ prepared by Alice. If 01 is the secret message, she has to choose $\sigma_z$ as the unitary operator. After performing the unitary operation on $A$, she will be ensued with $A'$.

Therefore,

$$\sigma_z(|\phi^+\rangle) = \sigma_z\left\{\frac{1}{\sqrt{2}}(|00\rangle + |11\rangle)\right\} = \frac{1}{\sqrt{2}}(|00\rangle - |11\rangle) = |\phi^-\rangle = A'.$$

**Step 2:** After receiving the operation result $A'$ from Alice, Bob will select one Bell state as an initial state according to his secret message and $A'$ by following the Table 2. Say, 11 is the secret message and $|\phi^-\rangle$ is the operation result of Alice, then the initial state of Bob will be $|\psi^+\rangle$. Now Bob will announce the same result $A'$ to Alice. Note that no need for Bob to encode the initial state using the unitary operator.

**Step 3:** If there is no eavesdropper, Alice will get the same result i.e. $A'$ from Bob. Alice will perform a measurement on the received operation result from Bob with $A'$. By doing so, if there is no eavesdropper she will be ensued with 1, else the measurement result will be 0. From this measurement result, Alice will check the presence of Eve. If there is an eavesdropper i.e. if the measurement result is 0, then Alice will announce to abort the communication.

Alice measurement

$$\langle A'|A'\rangle = \delta \Rightarrow \langle |\phi^-\rangle||\phi^-\rangle\rangle = 1$$

If the measurement result is $\delta = 1$, then Alice and Bob will be confirmed with the absence of eavesdropper, otherwise, they will abort the communication.

**Step 4:** For secure consideration, Alice and Bob will prepare a sufficient number of single particles $D$ in one of the four states $\{|0\rangle, |1\rangle, |+\rangle, |-\rangle\}$, where $|+\rangle = \frac{1}{\sqrt{2}}(|0\rangle + |1\rangle)$ and $|-\rangle = \frac{1}{\sqrt{2}}(|0\rangle - |1\rangle)$. By inserting $D$ to the initial state $A$ and $B$, it will become $A_D$ and $B_D$ respectively. Alice and Bob then exchange the new states $A_D$ and $B_D$ to each other simultaneously and execute the security checking on $D$. If error rate goes beyond the threshold, they will abort the communication. Otherwise, they will proceed to the next step. Upon getting the initial state $B$ and $A$, Alice and Bob can deduce the secret messages of each other according to Table 3.

In the example of our discussion, we have

$$(A', B) \rightarrow Msg_B \Rightarrow (|\phi^-\rangle, |\psi^+\rangle) \rightarrow 11$$

$$(A', A) \rightarrow Msg_A \Rightarrow (|\phi^-\rangle, |\phi^+\rangle) \rightarrow 01$$

Thus the messages are secretly exchanged between Alice and Bob without the help of any controller.

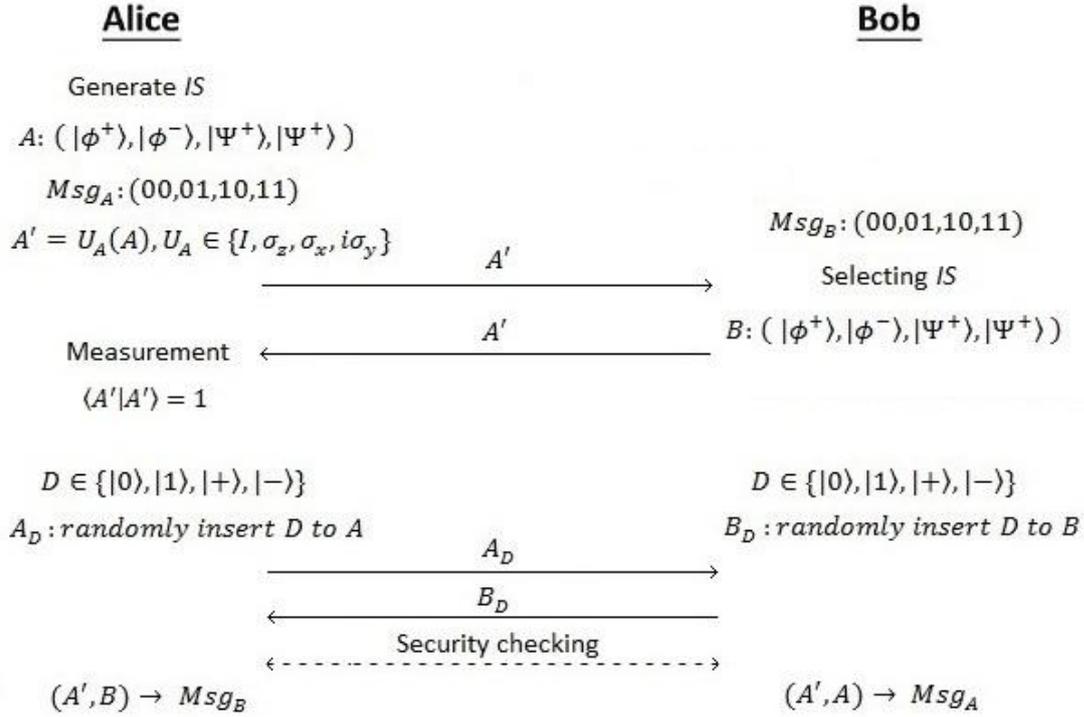

*Figure 2:* Controller independent BQDC protocol.

The presence of outsider can be realized by the measurement of Alice at the third step of the protocol. Once Alice's measurement will give the wrong measurement, Alice and Bob can abort the communication. Thus, the intercept-and-resend attack is also avoided in this protocol. Even if the outsider aware of the measurement results of Alice and Bob, he cannot do anything without knowing the initial states. Equivalently, all the four initial states are possible for any given measurement results of Alice and Bob (refer Table 3). Furthermore, no information is leaked out from either of the communicants. In C.H. Chang et al. protocol Alice and Bob doesn't know about the initial state generated by Charlie. If Charlie is not a genuine arbiter, he can distribute the wrong sequence among Alice and Bob. As a result, they will exchange the wrong message and they will aware of this at the last step of the protocol. Controller independent protocol does not experience such types of disadvantages.

*Table 3:* The relation between Alice's secret messages, initial states and Bob's secret messages, initial  states and their unitary operation results.

| Alice's secret messages \ Initial State (IS) → | $|\phi^+\rangle$ | $|\phi^-\rangle$ | $|\psi^+\rangle$ | $|\psi^-\rangle$ | Initial State (IS) ← / Bob's secret messages |
|---|---|---|---|---|---|
| 00 | $|\phi^+\rangle$ | $|\phi^-\rangle$ | $|\psi^+\rangle$ | $|\psi^-\rangle$ | 00 |
| 01 | $|\phi^-\rangle$ | $|\phi^+\rangle$ | $|\psi^-\rangle$ | $|\psi^+\rangle$ | 01 |
| 10 | $|\psi^+\rangle$ | $|\psi^-\rangle$ | $|\phi^+\rangle$ | $|\phi^-\rangle$ | 10 |
| 11 | $|\psi^-\rangle$ | $|\psi^+\rangle$ | $|\phi^-\rangle$ | $|\phi^+\rangle$ | 11 |

## 5 Conclusion

The first result of this work exhibits the importance of Bell states, which are being used as initial states in Chih-Hung Chang *et. al.* protocol. It is to be emphasized that maximally entangled states *only* can be used as initial states so that the communicants can choose their secret messages irrespective of the initial states generated by the controller. On the other hand, initial states can be chosen based on the secret messages, if the communicants are empowered with generating maximally entangled Bell states. In this case, the controller becomes insignificant.

    In proposed protocol, Alice and Bob are proficient of generating initial states according to their secret messages. This protocol does not require single particle encoding due to the adapted procedure. The intercept-and-resend attack is not possible at any step and information leakage between the communicants is also not possible in this proposal. Thus the proposed controller independent bidirectional quantum communication protocol is like a *conversion* between two persons without the help of any third person. At the same time, the level of security is also ensured in their conversion. In short, the controller independent bidirectional quantum direct communication protocol introduces a new scheme of quantum cryptography.